# Magnetic Behaviour of Assemblies of Interacting Cobalt-Carbide Nanoparticles

Pallabi Sutradhar [a], Shiv N. Khanna[b], and Jayasimha Atulasimha[a]

a. Dept. of Mechanical and Nuclear Engineering, Virginia Commonwealth Univ., Richmond, VA, USA

b. Dept. of Physics, Virginia Commonwealth Univ., Richmond, VA, USA

Email: sutradharp@vcu.edu (Pallabi Sutradhar) snkhanna@vcu.edu (Shiv N. Khanna), jatulasimha@vcu.edu (Jayasimha Atulasimha)

**Abstract:** Recent work [1] demonstrated high coercivity and magnetic moment in cobalt carbide nanoparticle assemblies and explained the high coercivity from first principles in terms of the high magnetocrystalline anisotropy of the cobalt carbide nanoparticles. In this work, we comprehensively model the interaction between the nanoparticles comprising the assembly and systematically understand the effect of particle size, distribution of the orientations of the nanoparticles' magnetocrystalline anisotropy axis with respect to the applied magnetic field, and dipole coupling between nanoparticles on the temperature dependent magnetic behavior of the nanoparticle assembly. We show that magnetocrystalline anisotropy alone is not enough to explain the large hysteresis over the 50K-400K temperature range and suggest that defects and inhomogeneties that pin the magnetization could also play a significant role on this temperature dependent magnetic behavior.

Permanent magnets are used in an increasing number of applications and are typically alloys that have 4f elements (rare earth materials). These 4f elements lead to both high magnetic moment and high magnetocrystalline anisotropy that lead to properties desirable in permanent magnets: high remanence and high coercivity [2]. However, the mining process for rare earth materials is detrimental to the environment, their yield is low and their extraction cost is high [3]. This motivates research on rare earth free permanent magnets with high coercivity and high magnetic moment.

One way to make such permanent magnets is by employing exchanged-coupled hard and soft magnetic material, which is known as 'exchange-spring magnet' [4]. Exchange spring magnets have high coercivity and high magnetic moment due to exchange interaction between a hard magnetic material (whose moment may not be very high) and a soft magnetic material with high magnetic moment. Thus, exchange spring magnets that have low rare earth material content (<15% hard magnetic layer) in combination with an exchange coupled high magnetic moment soft layer can exhibit the desirable properties of hard magnets [5]. Furthermore, by coupling an extremely hard FePt phase layer [6] with a soft high moment $Fe_3Pt$ phase layer, rare earth free hard magnets can be synthesized [7].

Another approach is the development of core-shell magnetic particles [8], [9]. The authors synthesized Co/CoO nanoparticles where exchange bias between the central soft ferromagnet Co with high magnetic moment and the surrounding antiferromagnetic oxide CoO greatly enhances the magnetic anisotropy of the core-shell system while retaining high magnetic moment. These nanoparticles have a blocking temperature close to room temperature 290K and coercivity of 0.59 T while exhibiting high saturation magnetic moment [8].

Recently, El-Gendy et al synthesized a phase of cobalt carbide ($Co_3C$) nanomagnets where the cobalt layers are far more separated than in bulk or thin film cobalt and embedded with intervening layers of the carbon atoms allowing only partial mixing between C and Co states. These magnets exhibit thermal and stable long range ferromagnetic order up to 573±2K (the blocking temperature). This high blocking temperature is due to the separate Co layers and mixing with carbon states which contributes to the unusually high magnetocrystalline anisotropy energy (MAE) [5] [10]. The observed unusually high MAE in $Co_3C$ was explained using first principles theoretical investigations [5].

Although DFT correctly predicted the role of structure and composition of $Co_3C$ on the high MAE and blocking temperature of individual nanoparticles, studies on the collective behavior of an assembly of nanoparticles are needed to understand the behavior of the synthesized nanomagnets. In this work, we have examined several unexplained aspects of the magnetic behavior of the $Co_3C$ nanoparticle assembly that have not been modeled or understood. These include effect of particle size, distribution in orientation of the nanoparticle crystal axis with the global magnetic field and dipolar interaction between the $Co_3C$ nanoparticles, which is investigated in this letter. Such an understanding is particularly important when one scales from understanding the behavior of individual $Co_3C$ nanoparticles to bulk materials produced from assemblies of such $Co_3C$ nanoparticles.

Our model for the magnetization behavior is described next, followed by a discussion on the results and a finally, a summary of the understanding developed from this work.

We consider a $Co_3C$ system with 16 particles dispersed on a flat surface as shown in Fig 1 a. In this system, each particle is assumed to act as a giant classical magnet due to the strong exchange coupling between the spins, a reasonable assumption for dimensions < 100 nm, (see Ref [11]). The particles are assumed to be spherical in shape with diameter ~4 nm. Such a system is shown in Fig 1. The variation of magnetization of the entire system with time under the influence of an effective field $H_{eff}$ is described by the Landau-Lifshitz-Gilbert (LLG) equation [12]:

$$(1+\alpha^2)\frac{d\vec{M}}{dt} = -\gamma \vec{M} \times \vec{H_{eff}} - \frac{\alpha\gamma}{M_s}\left[\vec{M} \times (\vec{M} \times \vec{H_{eff}})\right] \qquad (1).$$

Here $\gamma$ is gyromagnetic ratio and $\alpha$ is the damping factor.

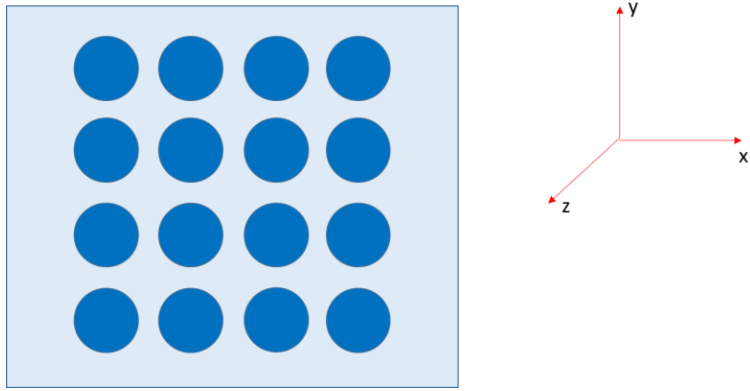

**Fig.1**: (a) 16 particle $Co_3C$ system, each particle with radius 2 nm and (b) the co-ordinate system.

The total effective field ($H_{eff}$) on each $Co_3C$ particle is the sum of fields due to Zeeman energy, magnetocrystalline anisotropy energy, dipole interaction energy and a random field that simulates the effect of thermal noise at a given temperature:

$$H_{eff} = H_{zeeman} + H_{magnetocrystalline\ anisotropy} + H_{dipole} + H_{thermal} \quad (2).$$

The field due to Zeeman energy is the applied magnetic field in the y-direction as shown in Fig. 1.

$$H_{zeeman} = H_{applied\ magnetic\ field}$$

The field due to magnetocrystalline anisotropy energy is given below:

$$H_{magnetocrystalline\ anisotropy} = \frac{2K_u}{\mu_0 M_{sat}} (\vec{u}.\vec{m})\ \vec{u} \quad (3).$$

Here $K_u$ is the magnetocrystalline anisotropy energy constant which is calculated by first principle DFT analysis ($K_u = 7.5 \pm 1 \times 10^5$ J/m³), $M_{sat}$ is saturation magnetization estimated from experiment [5], $\vec{u}$ is the direction of the easy axis for each $Co_3C$ particle, $\vec{m}$ is the magnetization direction and $\mu_o$ is the permeability of free space.

Consider two adjacent particles in a system (labeled as the $i^{th}$ and $j^{th}$ elements), whose magnetizations have polar and azimuthal angles of $\theta_i, \varphi_i$ and $\theta_j, \varphi_j$, respectively then magnetic dipole energy is given by:

$$E_{dipole}^{i-j} = \frac{\mu_0 M_s^2 \Omega^2}{4\pi R^3} \sum_{\substack{i-1 \\ j \neq i}}^{i+1} -2(\sin\theta_i \cos\varphi_i)(\sin\theta_j \cos\varphi_j) + (\sin\theta_i \sin\varphi_i)(\sin\theta_j \sin\varphi_j) + \cos\theta_i \cos\theta_j \quad (4).$$

where $M_s$ is the saturation magnetization, $\Omega$ is the volume of each $Co_3C$ particle and R is the separation between their centers.

The effective field due to dipole energy is:

$$H = -\frac{1}{\mu_0 \Omega} \frac{dE_{dipole}}{d\vec{M}} \quad (5).$$

Thermal noise at a given temperature manifests itself as a random magnetic field given by:

$$H_{thermal} = \sqrt{\frac{2k_b T \alpha}{\mu_0 M_s \gamma V h}} \qquad (6).$$

Here $k_b$ is Boltzman constant, T is the temperature and V is the volume of each $Co_3C$ particle.

After calculating the magnetization orientation as a function of time under an effective magnetic field using the LLG equation, we calculated the magnetization in the y-direction for each $Co_3C$ particle for a particular applied magnetic field (in the y-direction) using the equation given below:

$$m = \frac{1}{n}\sum_{i=1}^{n} \vec{M} \sin\theta_i \sin\phi_i \qquad (7).$$

In our study, we first calculated the response of the $Co_3C$ system without dipole interaction between $Co_3C$ particles by considering around 200 non-interacting Co particles whose magnetization was considered to evolve independently with time. In the next case we considered dipole interaction and chose a 16 particle system uniformly distributed on a surface as shown in Fig 1a. Here the magnetization of each particle was considered to evolve under the influence of dipole coupling of the neighboring particles. We set the separation between the surfaces of neighboring spherical Cobalt Carbide nanoparticles to be 0.1nm (separation distance assumed to be small to study the effect of large dipole coupling). While calculating the dipole interaction, we assumed that only the nearest neighbors have influence on the magnetization on each particle. Finally, we note that after applying each increment in magnetic field, we allowed sufficient time (typically ~10 to 100 nanoseconds) to allow the system to come to an equilibrium magnetic orientation (though there is always some fluctuation due to thermal noise), as we are interested in simulating quasistatic M-H curves.

We have simulated the effect of size of the particles, different distribution of the easy axis and effect of dipole interaction between particles.

1. **Effect of $Co_3C$ particle size on the overall response of the system:** Fig 2 shows the size effect of $Co_3C$ particles. The easy axes of these $Co_3C$ particles were distributed uniformly between 60° to 120° in the co-ordinate axis as shown in Fig 3b. In other words, the easy axis of the magnetic particles are uniformly distributed within ±30º from the direction of application of the magnetic field. The simulations were performed at a temperature of 50K. The figure shows that as we decrease the size of the $Co_3C$ particles the area of the hysteresis loop decreases as the effect of thermal noise at a given temperature increases when the particle size is reduced. If we continue to decrease size of the particles, they reach to superparamagnetic limit and the system would show no hysteresis. Fig. 2 shows that the particle size with radius 1.7 nm best matches the experiment at 50K. However, the experiments were performed with particles with a radius of around ~ 3nm [1]. The discrepancy between theory and experiment is potentially due to inactive outer carbon layers and it is likely that the magnetically active $Co_3C$ are have a nominal radius ~ 2 nm.

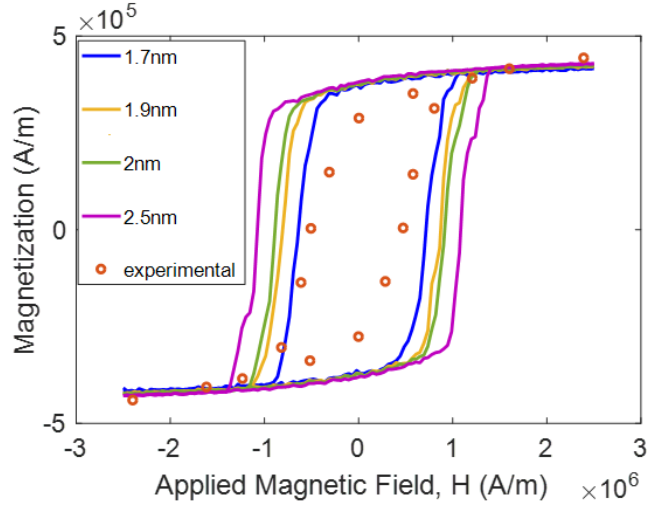

**Fig 2.** M-H curve for 1.7nm, 1.9nm, 2nm and 2.5nm radius $Co_3C$ particles. Experimental data from Ref [9].

1. **Effect of the distribution of easy-axes of $Co_3C$ particles for a specific size**: In Fig 3a we changed the distribution of easy axes while keeping the size of each particle 2nm and the temperature 50K. We can see from the figure that changing the distribution of the easy axes of the particles affects the shape of the hysteresis loop. When easy axes of all the particles are aligned along the applied magnetic field, the magnetization of all the particles rotate abruptly through 180º leading to a sudden change in magnetization. For the case of uniform distribution of the orientation of the easy axis with respect to the magnetic field direction, the easy axes which are aligned closer to the applied field switch abruptly but need smaller fields to drive the magnetization to saturation while the easy axes which are further away from the applied field rotate gradually but need larger fields to drive the magnetization to saturation. Therefore, there is no abrupt change in the magnetization for the uniformly distributed system as is shown in Fig 3a. By evaluating different kinds of distribution for our system we found out that the distribution for which the easy axes are distributed uniformly between 60° to 120° best matches the experiment. While no magnetic field was applied during the assembly of the $Co_3C$ nanoparticles (a strong magnetic field could potentially promote a preference for the magnetic easy axis of each particle aligning close to the direction of this applied field thus enhancing the coercivity and remanence of the assembly), the dipolar coupling between the these nanoparticles results in some texturing in the nanoparticle assembly. This is evident from simulations in Fig 3 (a) as a random distribution in orientation of the magnetic easy axis leads to a vastly different M-H curve than observed experimentally while magnetic nanoparticles' easy axes uniformly distributed within ±30º (i.e. 60° to 120°) from the direction of application of the magnetic field provides a better qualitative fit to the experimental data. Hence, in all other figures in this letter the simulations were performed assuming the magnetic nanoparticles' easy axes was uniformly distributed within ±30º from the direction of the magnetic field.

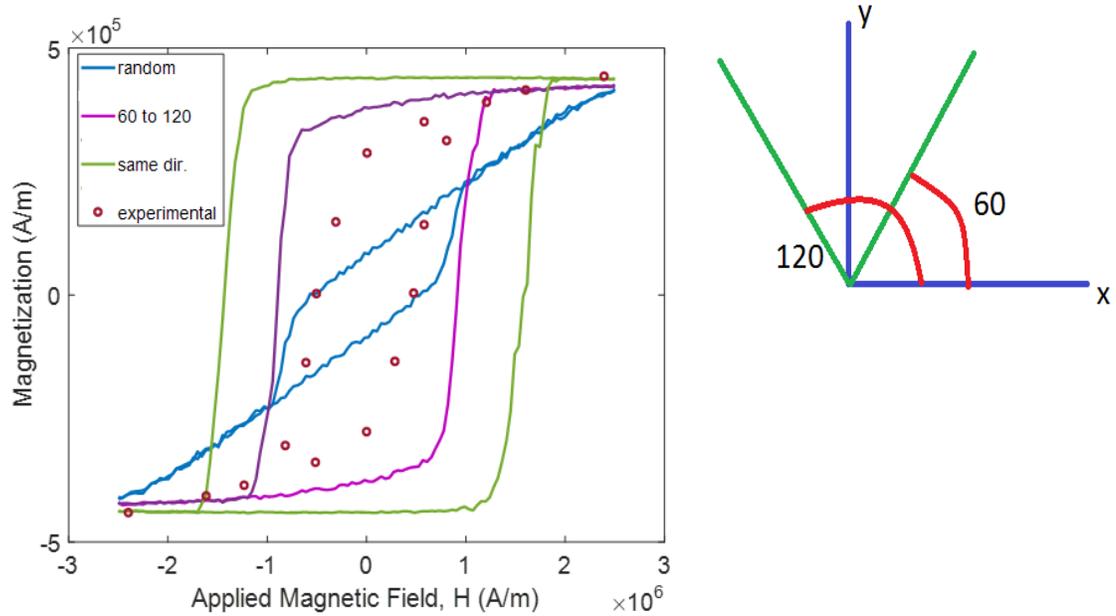

**Fig 3.** (a) M-H curve of $Co_3C$ system with easy axis (i) distributed randomly (ii) uniformly between 60° to 120° and (iii)all pointing in the same direction; compared with (iv)experimental data (b) Coordinate system showing the easy axis distribution between 60° and 120°. NOTE: Y-axis is the direction of application of the magnetic field. Experimental data from Ref [9].

2. **Effect of dipole interaction on the $Co_3C$ system**: In Fig. 4, we consider dipole interaction between $Co_3C$ particles while keeping the size of each particle 1.7nm, the temperature at 50K and the distribution of the easy axis between 60° to 120°. We can see from the figure that the area of the hysteresis loop with dipole interaction is slightly smaller than the loop with no dipole. Consider the assembly modeled in Fig 1, with magnetic field applied along the y-direction which orients the magnetization of each nanoparticle to eventually align parallel to it. The dipole coupling between nearest neighbors along the y-axis favor parallel orientation of their magnetization making it easier to rotate the magnetization to point along the y-axis, making the hysteresis loop smaller (compared with the simulations that do not consider dipole effect). However, the dipole coupling between nearest neighbors along the x-axis favor anti-parallel orientation of their magnetization, making it harder to fully magnetize the entire system, leading to slightly smaller net magnetization (compared to the simulations that do not consider dipole effect) at higher magnetic fields. It is interesting to note that the dipolar coupling has a relatively weak effect on the magnetization curves suggesting that the nanoparticles can be packed very closely to increase the saturation magnetization of the assembly without significantly degrading the coercivity.

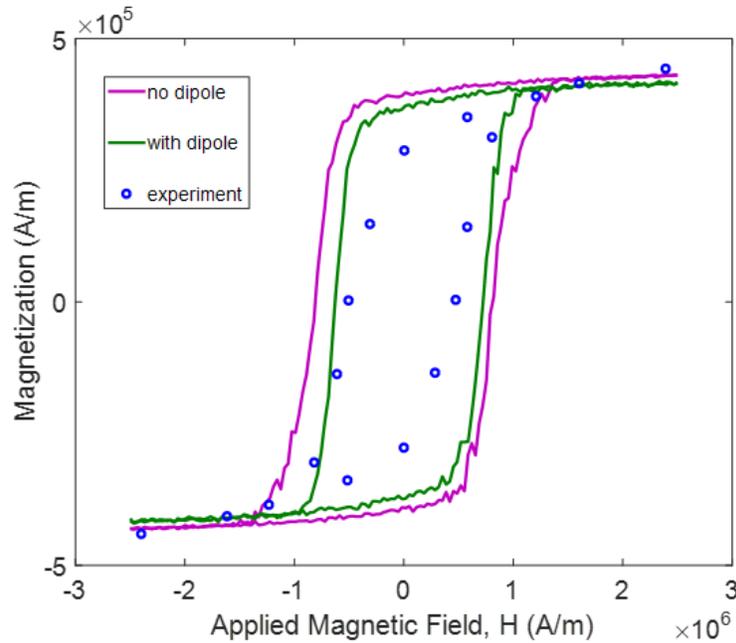

**Fig 4.** M-H curve of $Co_3C$ system with and without dipole interaction between the particles. Experimental data from Ref [9].

1. **Effect of temperature on the $Co_3C$ system**: In Fig 5, we simulate the effect of temperature on $Co_3C$ nanoparticles. For 1.7 nm particles, while the hysteresis is modeled well at 50K we get there is no hysteresis at 250K which differs significantly from experimental results. Magnetocrystalline Anisotropy Energy (MAE = $K_uV$, where $K_u$ is the magnetocrystalline anisotropy energy density and V is the volume of the nanoparticle) is proportional to the volume of the individual $Co_3C$ nanoparticles. From Fig 5b we can see that, at 250K, there is no hysteresis for 1.7 nm radius particles as the volume of these particles are small and therefore their anisotropy energy is small and comparable with the thermal energy at 250K. However, for particles with 2.3 nm radius at 250K we get a good match with the experimental hysteresis loop, though in this case the hysteresis at 50K would be over predicted (see Fig 5c). Further, Fig 5c compares the simulated M-H curves of the 2.3 nm radius nanoparticle at 50K, 250K and 400K with experimental data and shows that while hysteresis is correctly modeled at 250K, it is significantly over-predicted at 50K and slightly under predicted at 400K. This indicates that the magnetocrystalline anisotropy alone cannot explain the temperature dependence of the M-H curves. Defects and inhomogenoties that pin the magnetization can enhance the hysteresis over that is predicted from purely magnetocrystalline anisotropy considerations. Hence, it is likely that a smaller nanoparticle size (e.g. ~ 2 nm radius) may explain the hysteresis correctly at 50K but may not show a rapid decrease in hysteresis at higher temperatures and model the hysteresis correctly at 250K and 400K if the effect of magnetization pinning is correctly modeled. However, inclusion of such defects that pin the magnetization and modeling incoherent switching in such assemblies of nanoparticles in the presence of thermal noise and defects is beyond the scope of this work. But other research [13] suggests that such defects may be an important factor in modeling the behavior of magnetic nanoparticles and patterned devices.

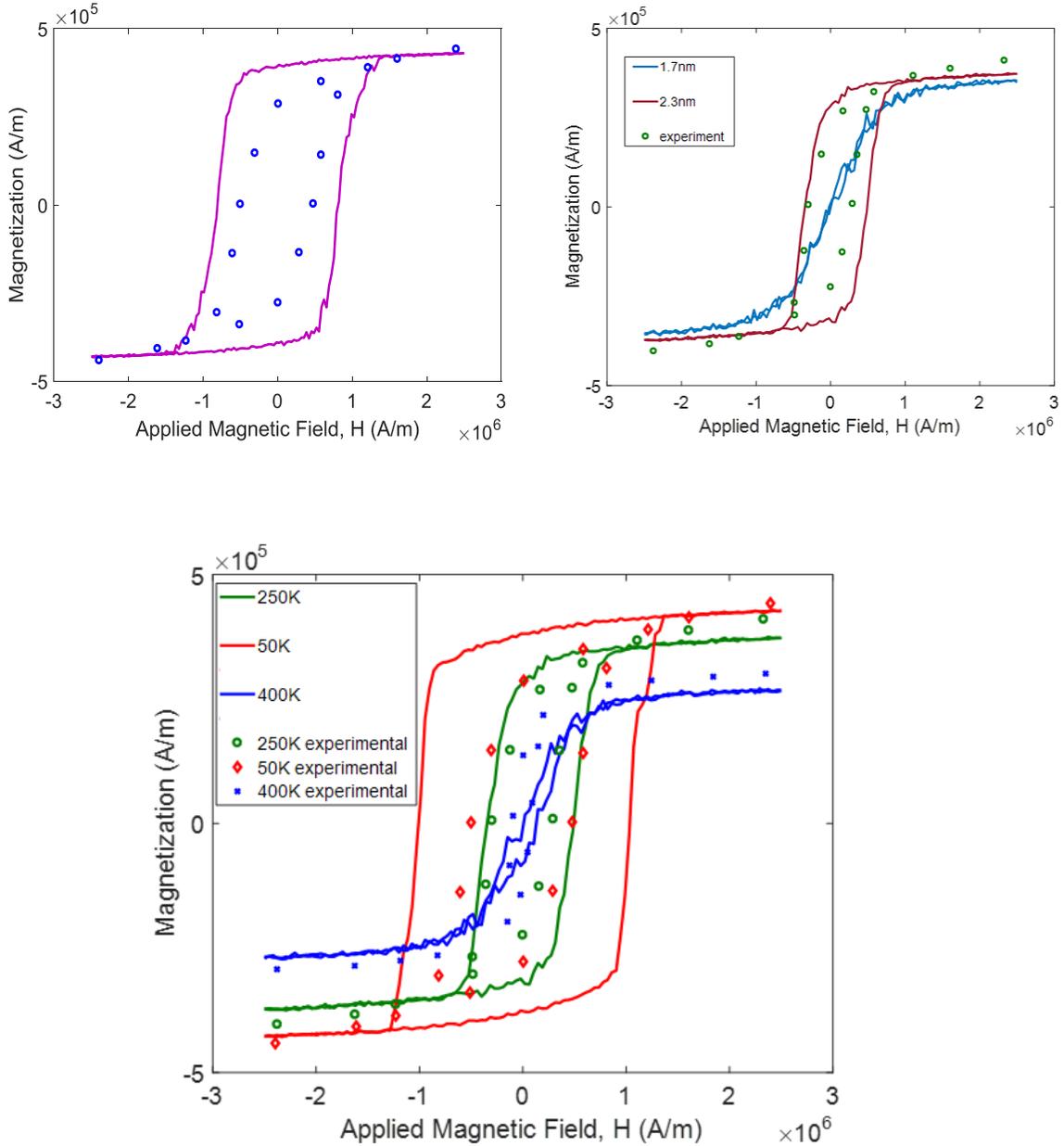

**Fig 5.** (a) M-H curve of 1.7nm $Co_3C$ particles system at 50K (b) M-H curve of 1.7nm and 2.3nm $Co_3C$ particles system at 250K. (c) Comparison of simulation vs. experimental data at 50K, 250K and 400K for particles of radius 2.3nm Experimental data from Ref [9].

We have explored the effect of nanoparticle size, orientation with respect to the magnetic field and dipolar coupling between these nanoparticles on the magnetic behavior of the $Co_3C$ nanoparticle assembly. While the magnetic behavior is extremely sensitive to particle size and orientation, the effect of dipole coupling is less significant. This suggests that one could pack nanoparticles with high densities in such assemblies to achieve high net magnetization without degrading the hard magnetic behavior (coercivity) significantly. Furthermore, particle size and distribution of orientation of the easy axis (of magneto crystalline anisotropy) can be used to optimize the magnetic properties of such assemblies. Finally, the temperature dependent behavior is qualitatively predicted and hysteresis is explained by magnetocrystalline anisotropy but the reduction in hysteresis with increasing temperature is significantly over predicted by the model. This suggests that magnetocrystalline anisotropy alone cannot describe the temperature dependent hysteresis. Defects that pin the magnetization could potentially enhance the hysteresis (and therefore coercivity) of these magnetic nanoparticles, making them retain favorable coercivity even at elevated temperatures.